\documentclass[prl,twocolumn,showpacs,amsmath,floatfix]{revtex4}

\usepackage{amssymb}
\usepackage{amstext}
\usepackage{amscd}
\usepackage{amsthm}
\usepackage{amsopn}
\usepackage{amsmath}
\usepackage{amsbsy}
\usepackage{indentfirst}
\usepackage{graphicx}

\newcommand{\be}{\begin{equation}}
\newcommand{\ee}{\end{equation}}
\newcommand{\bea}{\begin{eqnarray}}
\newcommand{\eea}{\end{eqnarray}}

\begin{document}

\author{Pankaj Mehta and Natan Andrei}

 \title{ Nonequilibrium Transport in Quantum Impurity Models \\
 (Bethe-Ansatz for open systems)}

\affiliation{Center for Materials Theory, Rutgers University,
Piscataway, NJ 08854}
 
\begin{abstract}

We develop an exact non-perturbative framework to compute steady-state
properties of quantum-impurities subject to a finite bias. We show
that the steady-state physics of these systems is captured by
nonequilibrium scattering eigenstates which satisfy an appropriate
Lippman-Schwinger equation.  Introducing a generalization of the
equilibrium Bethe-Ansatz - the Nonequilibrium Bethe-Ansatz (NEBA), we
explicitly construct the scattering eigenstates for the Interacting
Resonance Level model and derive exact, nonperturbative results for
the steady-state properties of the system.
 
\end{abstract}
\pacs{72.63.Kv, 72.15.Qm, 72.10.Fk   }
\maketitle

 The recent spectacular progress in nanotechnology has made it
  possible to study quantum impurities out-of-equilibrium
 \cite{GG}. The impurity is typically realized experimentally as a
  quantum dot, a tiny island of electron liquid attached via tunnel
  junctions to two leads (baths or reservoirs) held at different
  chemical potentials.  As a result of the potential difference, an
  electric current flows from one lead to another across the
  quantum impurity.  The description of such an out-of-equilibrium
  situation in a strongly correlated system is a long standing
  problem and has not been given even in the simplest case of when the
  system is in a steady state.

  In a steady state the system properties do not change with time even
  when out of equilibrium.  Such a state is reached only under
  special conditions: each lead needs to be a good thermal bath and
  infinite in size (equivalently, the bath level
  spacing tends to zero.)  It then follows that particles transferred
  from one lead to another dissipate their extra energy in the lead
  and equilibrate \cite{doyon}.

  There are two equivalent ways, time-dependent and time-independent,
 to describe the establishment of a steady state in the system.  In
 the time dependent picture the quantum impurity is coupled to the two
 baths in the far past, $t_0$, and is allowed to evolve
 adiabatically under the conditions described above.  After a
 sufficiently long time, at $t=0$ say, a steady state is reached. Two
 elements are required to fully determine the system: a hamiltonian to
 describe the time evolution and an initial condition, $\rho_0$,
 describing the system in the far past.  The hamiltonian is chosen to
 be of the form, $H(t)=H_0 + e^{\eta t}H_1$, where $H_0$ describes
 the two free leads (thermal baths), $H_1$ is the interaction term
 between the leads and the quantum impurity, and $\eta$ an
 infinitesimal parameter, small enough to ensure adiabaticity yet
 large compared to the level spacing in the leads. The initial
 condition is typically given by,
\bea
\rho_0 =
\frac{e^{-\beta(H_0-\sum_{i} \mu_i N_i)}}{Tr e^{-\beta(H_0-\sum_{i} \mu_i N_i)}} 
\eea
with $\mu_i$ and $N_i$ the chemical potential and number operator for
particles in lead $i$. Subsequently, at times $t \ge t_0$, the system
is described by a density matrix $\rho(t)= T\{ e^{i\int_{t_0}^t \,
dt^\prime H(t^\prime) } \}\rho_0 T\{ e^{-i\int_{t_0}^t \, dt^\prime 
H(t^\prime)
} \}$, and the properties of the system are calculated in the usual
manner, $ \langle \hat{O}(t) \rangle = Tr\{\rho(t) \hat{O}\}$. 
The establishment of a steady state follows, in this language,
from the existence of the limit $t_0 \to - \infty$ with the expectation value 
becoming time-independent, 
 $ \langle \hat{O} \rangle =  Tr\{\rho_s\hat{O}\}$ where $\rho_s=\rho(0)$. 

At $T=0$ the description simplifies. The initial condition is
typically given by a particular eigenstate of $H_0$,
$|\Phi\rangle_{baths}$, describing the baths, each with its own
chemical potential $\mu_i$. The steady state is then obtained by
evolving the initial state in time, $|\Psi\rangle_s = T\{
e^{i\int_{-\infty}^t \,dt^\prime H(t^\prime) } \}
|\Phi\rangle_{baths}$. The expectation values in the steady state are
computed from,
\bea
\langle \hat{O} \rangle = \frac{\langle \Psi|\hat{O}|\Psi\rangle_s}{\langle \Psi|\Psi\rangle_s}
\label{expectationvalue0}
\eea

 An equivalent way to describe a non-equilibrium steady-state is by
 means of a time-independent scattering formalism. The state
 $|\Psi\rangle_s$ is obtained as an eigenstate of the full hamiltonian
 $H= H_0 + H_1$, satisfying the Lippman-Schwinger equation,
\bea 
|\Psi\rangle_s
= |\Phi\rangle_{baths} + \frac{1}{E-H_0\pm i\eta }H_1|\Psi\rangle_s 
\eea
with $|\Phi\rangle_{baths}$ - the incoming state.  The scattering
eigenstate $|\Psi\rangle_s$ can be viewed as consisting of 
incoming particles (the two free Fermi-seas) described by 
$|\Phi \rangle_{baths}$ and reflected  outgoing particles 
given by the second term in the above equation. Once again 
two elements are required to
fully determine the system: a hamiltonian and a boundary condition,
$|\Phi\rangle_{baths}$, which describes the scattering state far from the
impurity. Note that previously, in the time-dependent picture
$|\Phi\rangle_{baths}$ played the role of an initial condition rather
than a boundary condition.  The finite temperature description in this
formalism is obtained by summing over scattering states weighted
according to the Boltzman weights of the corresponding incoming
states.

The construction of such eigenstates is a formidable task in general.
We shall show, however, that it can be carried out for a class of
integrable impurity models that includes the Interacting Resonance
Level Model (IRLM) and the Kondo Model. The Bethe
Ansatz solution of these integrable models in equilibrium  has
led to a full understanding of their thermodynamic properties. It is
based on solving the hamiltonian of a closed system, typically with
periodic boundary conditions.  We shall present in this letter a
significant generalization of the Bethe Ansatz approach to {\it open
systems} with boundary conditions imposed by the leads.  This
approach, the Non-equilibrium Bethe-Ansatz (NEBA), allows  us to
construct the fully interacting multi-particle scattering eigenestates  
and  compute non-equilibrium transport properties, extending Landauer's
original approach \cite{land}. 
We remark here that our approach differs significantly
from the recent interesting work by Konik et al.
\cite{konik} who also used integrability to compute transport. In
contrast to their work we model the leads as  free  Fermi seas rather 
than coupling the chemical potentials to dressed excitations.

We focus on the IRLM at $T=0$ and defer treatment of other models to
later publications. The IRLM, $
 H_{\rm{IRL}}= \sum_{i=1,2,\vec{k}} \epsilon_k \psi_{i \vec{k}}^\dagger
\psi_{i \vec{k}} +  \epsilon_d d^\dagger d 
+\;\frac{t}{\sqrt{2}}\sum_{i=1,2, \vec{k}} ( \psi_{\vec{k}}
^\dagger d + h.c.) +\; 2U\sum_{i=1,2,\vec{k},\vec{k'}} \psi_{i \vec{k}}^\dagger\psi_ {i \vec{k'}}d^\dagger d 
$, describes a resonant level, $\epsilon_d
d^\dagger d$, coupled to two baths of spinless electrons via tunneling
junctions with strength $t$. There is also a Coulomb interaction $U$
between the the level and the baths. 
The model is closely related to the anisotropic Kondo model \cite{WF},
 with the charge states $n_d=0,1$ playing the role of spin states, and $\epsilon_d$ playing the role of a local magnetic field.

Performing some standard manipulations for impurity models: expanding
in angular modes around the impurity, keeping only the s-modes,
unfolding the model, and linearizing around the two Fermi points we have,
\bea 
H_{\rm{IRL}}&=&  -i \sum_{i=1,2} 
\int \, dx \, \psi_i^\dagger(x) \partial \psi_i(x) + \epsilon_d d^\dagger d  \\
&+& \frac{t}{\sqrt{2}}(\sum_{i=1,2} \psi_i^\dagger (0) d 
+h.c.)+ 2U\sum_{i=1,2}\psi_i^\dagger(0)\psi_i (0) d^\dagger d. \nonumber
\label{RLMHamiltonian}
\eea
The model thus obtained is a renormalizable field theory which
requires introduction of a cut-off procedure to render it finite. The
values of the bare parameters $U, \epsilon_d, t$ will be renormalized
as the cut-off is removed to yield a physical theory. The renormalized
theory captures the universal physics - where voltages and
temperatures are small compared to the cut-off (bandwidth $D$).  The
chemical potentials for the leads are not included in the
Hamiltonian. Instead, they enter as nonequilibrium boundary conditions
specifying the scattering-state far from the impurity.

We wish to calculate the  expectation values in the steady state of the dot
occupation, $\hat{n}_d=d^\dagger d$, and the current operator, $\hat{I} =
\frac{i}{2}\sum_{j=1,2} (-1)^j t /\sqrt{2}(\psi_j^\dagger(0)d - h.c)$, the
latter deduced from  $ \hat{I}= \frac{1}{2}[(N_1-N_2), H]$.

To construct the scattering states we shall use a new Bethe-Ansatz
technique which, unlike the traditional approach based on closed
systems and periodic boundary conditions, allows the determination of
a state by boundary conditions imposed asymptotically. We shall build
the many body scattering state using single particle scattering states 
that incorporate the boundary conditions. It is convenient to introduce the
symmetric/anti-symmetric basis defined by $ \psi_{e/o}(x) =
\frac{1}{\sqrt{2}}(\psi_1(x) \pm \psi_2(x))$, in terms of which the
hamiltonian separates into even and odd parts, $ H_e = -i\int \, dx \,
\psi_e^\dagger(x) \partial \psi_e(x)
+U\psi_e^\dagger(0)\psi_e^\dagger(0) d^\dagger d + t( \psi_e^\dagger
(0) d +h.c.)+ \epsilon_d d^\dagger d $, and $ H_o = -i\int \, dx \,
\psi_o^\dagger(x) \partial \psi_o(x)
+U\psi_o^\dagger(0)\psi_o^\dagger(0) d^\dagger d. $ The boundary
conditions, however, are imposed in the physical basis, $\psi_{1/2}$,
requiring appropriate combinations of both the even and odd
sectors. Both hamiltonians conserve the number of particles: $H_e$
commutes with $ \bar{N}_e= N_e + N_d$ and $H_o$ with $N_o$. The
construction proceeds by considering the $N$-particle sacttering
solutions, $N=1,2, \cdots$ It is important to note that in doing so we
have imposed cut-offs on the theory. Only upon taking the limit $N, L
\to \infty$ is the field theory regained and the results become
universal.

 The single-particle eigenstates
of the model take the form,
 \be
 \int dx \, [ A
(g_p(x)\psi_e^\dagger(x)+ e_p d^\dagger) +B h_p(x)\psi_o^\dagger(x)
]|0 \rangle
\label{generalsingleparticle}
\ee 
with $|0 \rangle$ the empty vacuum and $A$ and $B$ arbitrary constants
chosen to satisfy the nonequilibrium boundary conditions. We are
interested in two solutions, labeled $\pm$, to the Schrodinger
equation for these eigenstates,
\bea 
g_p(x) &=& \frac{2e^{ipx}}{1+e^{i\delta_p}} \left[\theta(-x) +
e^{i\delta_p}\theta(x)\right],\quad (g_p(0) =1) \nonumber\\ h_p^\pm(x)
&=& \frac{2e^{ipx}}{1+e^{i\delta_p}} \qquad \qquad \qquad \qquad \quad
\qquad \;\; x \neq 0 \nonumber \\ h_p^\pm(0) &=& \pm
\frac{(p-\epsilon_d)e_p e^{ipx}}{t}= \pm g_p(0)e^{ipx} \qquad  x =
0 \label{oneparticlesolution} 
\eea
 with $ e_p = tg_p(0)/(p-\epsilon_d)
$ and $ \delta_p = 2\arctan \left[\frac{t^2}{2(p-\epsilon_d)}
\right]$. In writing these states, we have chosen the regularization
scheme: $\theta (\pm x)\delta ( \pm x) = \frac{1}{2} \delta$.
Note, that we take $h_p(x)$ to be discontinuous at zero.  This
unorthodox choice of solution is allowed by the linear derivative.
Theories with linear derivatives -- as realized by Dirac long ago --
are implicitly many-body theories. To calculate physical observables,
one must first fill the Fermi-sea from a lower cut-off $D$ to the
Fermi energy. Since the universal many body physics is only sensitive
to the amplitude of $h_p(x)$ before and after the impurity,
renormalizability implies that physical observables are insensitive
to the  choice of
discontinuities in $h_p(x)$ \cite{AFL}.

 We construct two kinds of single-particle
scattering states, namely those with incoming particles from lead-1,
$|1p \rangle$, and those with incoming particles from lead-2, $|2p
\rangle$, with $p$ the momentum of the incoming particle. Choosing
$A=B$ in eq(\ref{generalsingleparticle}), the amplitude for an
incoming particle from lead-2 vanishes and we get,
\bea 
|1p\rangle &=& \int dx\,
e^{ipx}\left[\frac{2}{1+e^{i\delta_p}}\left([2\theta(-x) +
(e^{i\delta_p}+1)\theta(x)]\psi_1^\dagger \right. \right. \nonumber \\
&+& \left. \left. [(e^{i\delta_p}-1)\theta(x)]\psi_2^\dagger\right) +
\sqrt{2} e_pd^\dagger \delta(x) \right] |0 \rangle \nonumber \eea
Conversely, choosing $A=-B$, the amplitude for an incoming particle
from lead-1 vanishes and we get the state $|2p \rangle$, given by the
above expression with $\psi_1^\dagger(x)$ and $\psi_2^\dagger(x)$
interchanged.  It is convenient to introduce the operators, $
\alpha_{1/2p}^\dagger(x)= g_p(x) \psi_e^\dagger(x) \pm h_p^{\pm}(x)
\psi_o^\dagger(x) + e_p \delta(x)d^\dagger$, in terms of which the
scattering states are $|1/2p\rangle =\int dx e^{ipx}
\alpha_{1/2p}^\dagger(x)|0\rangle $. The single particle scattering
eigenstates are depicted in FIG. 1.
\begin{figure}[t]
\includegraphics[width=1.0\columnwidth, clip]{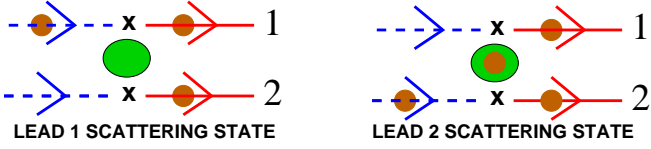}
\caption{
\label{fig:singlescattering}
There are two types of single particle scattering states. In type-1
scattering state, an incoming electron in lead 1 is scattered by
the impurity and can hop on to either lead 1 or lead 2. Notice there
is no amplitude for an electron to initially be in lead 2. In a type-2
scattering state, the role of the two leads is reversed. }
\end{figure}

The most general two-particle eigenfunction is of the form, 
\bea 
&& \int \int
 [Ag(x_1,x_2)\psi_e^\dagger(x_1)\psi_e^\dagger(x_2) +C h(x_1,
x_2)\psi_o^\dagger(x_1)\psi_o^\dagger(x_2)  \nonumber\\ 
&& + \int \int Bj(x_1,x_2)\psi_e^\dagger(x_1)\psi_o^\dagger(x_2) ] |0\rangle \nonumber \\
&& + \int [Ae(x) \psi_e^\dagger(x)d^\dagger 
+
Bf(x)\psi_0^\dagger(x)d^\dagger] |0\rangle
\label{2part}
\eea
with, 
\bea
2g(x_1,x_2) &=& 
g_p(x_1)g_k(x_2)Z(x_1-x_2)-
(1 \leftrightarrow 2) \nonumber \\ 
2h(x_1,x_2)&=& h_p(x_1)h_k(x_2)Z(x_1-x_2)-
(1\leftrightarrow 2) \nonumber \\
j^{ab}(x_1,x_2) &=&g_p(x_1)h_k^a(x_2)Z(x_1-x_2) \nonumber \\
&+&(-1)^{ab}
g_k(x_1)h_p^b(x_2)Z(x_2-x_1) 
\eea
with $a, b =\pm$, $g_p(x)$ and $h_p^{a/b}(x)$ being
the single particle eignefunctions eq(\ref{oneparticlesolution}) and
 $ Z(x_1-x_2) = e^{i\varphi(p,k)\rm{sgn}(x_1-x_2)}$ with $
e^{2i\varphi(p,k)} = (i+ \frac{U}{2}\frac{p-k}{k+p-2\epsilon_d})/( i-
\frac{U}{2}\frac{p-k}{k+p-2\epsilon_d})$. The constants $A$,
$B$, and $C$ are determined by the nonequilibrium boundary conditions. 
In this solution, we made use of 
freedom afforded by the linear dispersion  to choose the
two-particle S-matrix between all electrons to be the same.  This 
allows us to easily generalize the construction to $N$ particle wave functions
 yielding the
fully-interacting scattering state,
\bea
|\Psi \rangle_s =\int dx  \Psi(x_1 \cdots x_N)
 \prod_{u=1}^{N_1}\alpha_{1p_u}^\dagger(x_u)\prod_{v=N_1+1}^{N_1+N_2}\alpha_{2p_v}^\dagger(x_v)|0 \rangle \nonumber \\
 \Psi_s(x_1 \cdots x_N)=
e^{i\sum_j p_j x_j} e^{i\sum_{s<t} \varphi(p_s,p_t)sgn(x_s-x_t)} \quad \qquad  \quad
\label{finiteUscatstate}
\eea

Recall that for $|\Psi \rangle_s$ to describe a non-equilibrium
steady-state the incoming particles in the region $\{x_j\} \le 0$ must
be described by $|\Phi\rangle_{baths}$. In the coventional Fock basis
$|\Phi\rangle_{baths}$ is given by: $ \prod_{u=1}^{N_1} e^{i\sum_u
\bar{p}_u x_u }\prod_{v=N_1+1}^{N_1+N_2} e^{i\sum_v \bar{p}_v x_v}$, with the
Fock momenta $\{\bar{p}_j \}$ satisfying: $ -D \le \bar{p}_u \le
\mu_1$ and $ -D \le \bar{p}_v \le \mu_2$.  Notice, however, that in
$\Psi_s(x_1 \cdots x_N) $ there is a two particle $S$-matrix, $S=e^{2i
\varphi(p,k)}$, between incoming particles in each lead though the
particles are free electrons. The presence of this non-trivial
S-matrix forces a choice of a different, ``Bethe-Ansatz'', basis of
eigenstates for the free Fermi seas in the leads, inherited from the
interacting model when the coupling to the impurity is turned off
\cite{natan}. In order impose the boundary condition in the
Bethe-Ansatz basis, the incoming particles ``Bethe-Ansatz'' momenta
$\{ p_j \}$ in $|\Psi \rangle_s$, thus far undetermined, must be
appropriately chosen. This is done below by solving ``free-field''
Bethe-Ansatz equations.

The steady-state current and dot occupation in the non-equilibrium
steady-state is computed from eq(\ref{expectationvalue0}) with
$\hat{O}$ the appropriate operator and $|\psi \rangle$ given by
eq(\ref{U=0scatteringstate}). When computing this expectation value,
one must take the system size, $L$, to infinity (recall that no steady
state can be reached otherwise). In this limit, scattering states
of type 1 and 2 are orthogonal and we find,
 \bea
\langle I \rangle_s = \sum_{u=1}^{N_1} \frac{\Delta^2}{(p_u-\epsilon_d)^2 +
\Delta^2} - \sum_{v= N_1 +1}^{N_1+N_2}
\frac{\Delta^2}{(p_{v}-\epsilon_d)^2 + \Delta^2} \nonumber \\
\langle n_d \rangle_s = \sum_{u=1}^{N_1} \frac{\Delta}{(p_u-\epsilon_d)^2 +
\Delta^2} + \sum_{v= N_1 +1}^{N_1+ N_2}
\frac{\Delta}{(p_{v}-\epsilon_d)^2 + \Delta^2}. \nonumber \eea
with $\Delta= t^2/2$. In the non-interacting case, $U=0$, imposing the
boundary condition in the thermodynamic limit, the sums are replaced
by integrals over $\rho_i$ - product of the density of states
$\nu=1/2\pi$ and the Fermi-Dirac function - that describe the
distribution of momenta in each lead, e.g. at $T=0$, $\rho_i(p) =
\frac{1}{2 \pi} \theta(k^i_o -p)$, with $k^i_o$ set by $\mu^i$. We
obtain the standard RL results \cite{meir},
\bea
\langle I \rangle_s &=& \int \, dp \left[ \rho_1(p) - \rho_2(p) \right]
\frac{\Delta^2}{(p-\epsilon_d)^2 + \Delta^2} \nonumber \\
\langle n_d \rangle _s &=& \int \, dp \left[\rho_1(p) + \rho_2(p)\right]
\frac{\Delta}{(p-\epsilon_d)^2 + \Delta^2}
\label{steadystateexpectation}
\eea

In the interacting case, however, $\rho_i(p)$ are no longer
Fermi-Dirac distributions. As explained above, the presence of the
non-trivial $S$-matrix requires the distribution in each lead to be
obtained by solving a set a free Bethe-Ansatz equations.  For
$T=0$, the distributions, $\rho_i(p), i=1,2$ satisfy,
 \bea
\rho_1(p)= \frac{1}{2\pi}\theta( k^1_o-p) - \sum_{j=1,2}
\int_{-\infty}^{k^j_o}  {\cal K}(p,k)  \rho_j(k)\; dk
 \nonumber \\
\rho_2(p)= \frac{1}{2\pi}\theta( k^2_o-p) - \sum_{j=1,2}
\int_{-\infty}^{k^j_o}  {\cal K}(p,k)  \rho_j(k)\;dk \nonumber
\eea
with $k^i_o, i=1,2$  upper bounds on the distributions of $k$ set by the
chemical potentials $\mu_i$ (we choose $k^1_o > k^2_o$), ${\cal K}(p,k) = \frac{U}{\pi} (k-\tilde{\epsilon}_d)/
[(p+k-2\tilde{\epsilon}_d)^2+\frac{U^2}{4}(p-k)^2]$ and $\tilde{\epsilon}_d=\epsilon_d
-U\sum_{i=1,2}\frac{N_i}{L}$. The equations need be solved in the presence of a cut-off $D$, $-D \le k$.
For $T>0$ one
needs to solve the corresponding finite temperature Thermodynamic
Bethe Anatz  equations.

In the $U=\infty$ limit, these equations can be solved by a standard, if tedious,Wiener-Hopf method yielding the results
$\langle I \rangle_s = \frac{\Delta}{2\pi}
\left(\frac{T_k}{\Delta} \right)\left[\tan^{-1}\frac{\mu_1-\epsilon_d}{T_k}
-\tan^{-1}\frac{\mu_2-\epsilon_d}{T_k}\right]$ and
$\langle n_d \rangle_s=\frac{1}{2}+ \frac{1}{2
\pi}\left(\frac{T_k}{\Delta}\right)\left[\tan^{-1}\frac{\mu_1-\epsilon_d}{T_k}
+\tan^{-1}\frac{\mu_2-\epsilon_d}{T_k}\right]$
where $T_k = D\left(\frac{\Delta}{D}\right)^{\frac{2\pi}{\pi +\zeta}}$
with $e^{i\zeta(U)} = \frac{(1-\left[\frac{U}{2}\right]^2+
2i\frac{U}{2}}{1 + \left[\frac{U}{2}\right]^2}$. $T_k$ is a new low
energy scale in the problem, related to the Kondo temperature in the
anisotropic Kondo model. It  is held fixed as the
cut-off and  $U$ are  sent to infinity.

More generally, these equations
must be solved numerically with the bandwidth, $D$, much larger than
all parameters in the problem to ensure we are in a universal regime. 
\begin{figure}[t]
\includegraphics[width=0.9\columnwidth, clip]{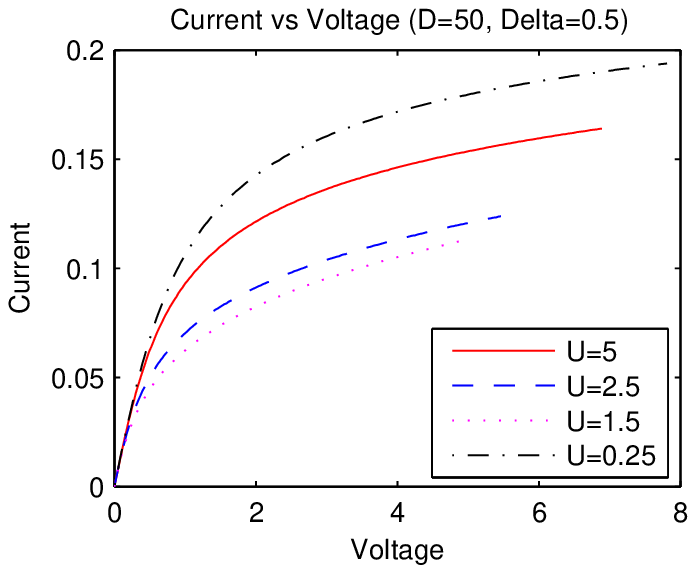}
\includegraphics[width=0.9\columnwidth, clip]{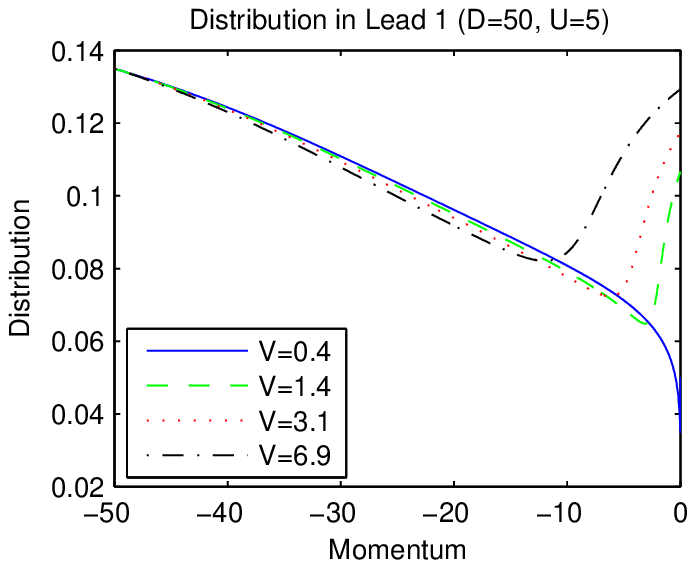} 
\caption{
\label{fig:singlescattering}
Here we show the current as a function of voltage for various $U$  
with $\mu_{1/2}=\tilde{\epsilon_d} \pm V/2$ for fixed bandwidth $D$
and $\Delta$. Note,the current is not monotonic in $U$.
We also show the distribution in lead 1 as a function of momentum for
various voltages where with out loss of generality  we take $k_o^1=0$.}
\end{figure}
In FIG. 2, we plot our results for the current as a function of voltage for 
various values of $U$. Notice the current is non-monotonic in U with
a duality between small and large $U$. This duality holds for all
$\Delta$. We also plot the distribution function in lead 1,
$\rho_1$, as a function of momentum for various voltages. The strong
dependence on momentum and voltage of these distributions  is a hallmark
of the nonequilibrium physics.

For $\mu_1 = \mu_2$ the system reverts to equilibrium and our
  construction can be compared with traditional Bethe Ansatz
  approaches.  There is no current in this case, and the dot occupation
  can be obtained from the impurity energy, given at zero temperature
  by $ E_{imp} = \int \, dp \rho (p) \delta_p$ with $\rho(p)$
  determined by the TBA equation. Hence, ignoring
  $\partial_{\epsilon_d} \rho(p)$, which is suppressed by $N/L$, we have $
  \langle n\rangle_d = \partial_{\epsilon_d} E_{imp} = \int \, dp \rho
  (p) \partial_{\epsilon_d} \delta_p$. Since $\partial_{\epsilon_d}
  \delta_p = 2\Delta/((p-\epsilon_d)^2 + \Delta^2)$, it coincides with
  eq(\ref{steadystateexpectation}) when $\rho_1(p) =\rho_2(p) =
  \rho(p)$.

In conclusion, we have presented an exact solution of a strongly
correlated impurity model out of equilibrium. The solution is given in
terms of the scatteringing states that characterize the nonequilibrium
steady state.  The generalization to finite temperature or to more
than two leads is straightforward. The latter allows the computation
the nonequilibrium density of states \cite{eran} which is of
experimental interest. We believe the framework we introduced is very general
and can be applied to most integrable models. Thus far we have
constructed current carrying scattering states for the Anderson and
Kondo models, though we do not know a general criterion for the
framework's applicability. Neither do we have a classification of the
operators whose non-equilibrium expectation values can be calculated.

{\bf Acknowledgments}: This work was started in collaboration with
Y. Gefen during a visit to the Weizmann Institute in Feb. 2001. We
thank him for his warm hospitality and generous help. We are grateful
to C. Bolech, E. Boulat, O. Parcollet, A. Rosch and especially to
B. Doyon and A. Schiller for numerous useful and enlightening
discussions. The authors were supported in part by BSF grant 4-21388.

\end{document}